\documentclass[a4paper]{article}

\usepackage[utf8]{inputenc}

\usepackage{multicol}
\usepackage{amsthm}
\usepackage{bm}
\usepackage{physics}

\usepackage[dvipdfmx]{graphicx}

\usepackage{bmpsize}
\usepackage{amssymb}
\usepackage{amsmath}
\usepackage{color}
\usepackage{amsthm}
\usepackage{comment}

\newcommand{\BA}{\begin{eqnarray}}
\newcommand{\EA}{\end{eqnarray}}

\definecolor{dgreen}{rgb}{0.0, 0.5, 0.0}

\begin{document}

\fontsize{14pt}{16.5pt}\selectfont

\begin{center}
\bf{
Dynamical properties of discrete negative feedback models
}
\end{center}
\fontsize{12pt}{11pt}\selectfont
\begin{center}
Shousuke Ohmori$^{1,2 *)}$ and Yoshihiro Yamazaki$^3$\\ 
\end{center}

\noindent
$^1$\it{National Institute of Technology, Gunma College, Maebashi-shi, Gunma 371-8530, Japan}\\
$^2$\it{Waseda Research Institute for Science and Engineering, Waseda University}{Shinjuku, Tokyo 169-8555, Japan}\\
$^3$\it{Department of Physics, Waseda University, Shinjuku, Tokyo 169-8555, Japan}\\

\noindent
*corresponding author: 42261timemachine@ruri.waseda.jp\\
~~\\
\rm
\fontsize{11pt}{14pt}\selectfont\noindent

\baselineskip 30pt

\noindent
{\bf Abstract}\\
%
%
Dynamical properties of tropically discretized and max-plus negative feedback models are investigated.
Reviewing the previous study [S. Gibo and H. Ito,  J. Theor. Biol. 378, 89 (2015)], the conditions under which the Neimark-Sacker bifurcation occurs are rederived with a different approach from their previous one.
Furthermore, for limit cycles of the tropically discretized model, 
it is found that ultradiscrete state emerges 
when the time interval in the model becomes large.
For the max-plus model, we find the two limit cycles; one is stable and the other is unstable.
The dynamical properties of these limit cycles can be characterized by using the Poincar\'{e} map method.
Relationship between ultradiscrete limit cycle states for the tropically discretized and the max-plus models is also discussed.

\bigskip

\bigskip




%
%

\section{Introduction}
When a continuous dynamical system is discretized, the discretized dynamical system can exhibit dynamical behaviors that the original continuous system never shows as in the well-known case of the logistic system 
for population of biological individuals\cite{May1976}.
Which model to apply, continuous or discrete, is determined by real phenomena focused on.
Negative feedback model in biological systems is also the similar case, 
and its continuous equation is given as 
\begin{equation}
    \frac{dx}{dt} = y-x,
    \;\;\;\;
    \frac{dy}{dt} = \frac{1}{1+x^m}-\frac{y}{b}, 
    \label{eqn:00a} 
\end{equation}
where $x=x(t),y=y(t)>0$ and $b,m$ are positive parameters\cite{Griffith1968}.
$m$ is called the Hill coefficient. 
For this continuous negative feedback model, it has been confirmed that 
there is no limit cycle solution\cite{Griffith1968,Kurosawa2002}.
On the other hand, based on eq.(\ref{eqn:00a}), the following discretized negative feedback model has been derived\cite{Gibo2015}:  
\begin{equation}
    x_{n+1} = \frac{x_n+\tau y_n}{1+\tau},
    \;\;\;\;
    y_{n+1} = \frac{y_n+\frac{\tau}{1+x_n^m}}{1+\frac{\tau}{b}}, 
    \label{eqn:trop1} 
\end{equation} 
where $x_{n}=x(n\tau)$, $y_{n}=y(n\tau)$, $n = 0, 1, 2, \ldots$, and 
$\tau>0$ corresponds to the time interval.
Gibo and Ito showed that the discretized model, eq.(\ref{eqn:trop1}), exhibits Neimark-Sacker bifurcation and has limit cycles solutions\cite{Gibo2015}.
For obtaining eq.(\ref{eqn:trop1}),  
they applied the tropical discretization\cite{Murata2013,Matsuya2015} 
to eq.(\ref{eqn:00a}).
They obtained conditions 
under which the Neimark-Sacker bifurcation occurs and 
the limit cycle solutions emerge in eq.(\ref{eqn:trop1}).
Furthermore, they derived the max-plus negative feedback model 
by ultradiscretization\cite{Tokihiro2004} 
and numerically showed existence of oscillatory solutions.  
They argued that the negative feedback model with discrete time steps is appropriate for biochemical situations where the rate of degradation is lower than that of synthesis and the threshold for feedback regulation is small. 
Moreover, even if $\tau$ is infinity, the tropically discretized model is considered to be still valid for systems where the successive reactions take place at prolonged intervals in biochemical processes.
Then it is meaningful to understand dynamical properties of eq.(\ref{eqn:trop1}) for application to real phenomena.

In this letter, by adopting our approach for identifying 
the types and stability of fixed points 
in tropically discretized dynamical systems\cite{Ohmori2023}, 
we review the previous study by Gibo and Ito\cite{Gibo2015}.
After that, we report results of further investigation 
for dynamical properties of the tropically discretized 
and the max-plus negative feedback models.
%

\section{Dynamical properties of eq.(\ref{eqn:trop1})}

First, we apply our systematic approach\cite{Ohmori2023} 
to eq.(\ref{eqn:trop1}), which is formally in the form of the following set of equations:
\begin{equation}
    x_{n+1} = x_{n} \frac{x_{n}+\tau f_{1}(x_{n}, y_{n})}{x_{n}+\tau g_{1}(x_{n}, y_{n})},\;
    y_{n+1} = y_{n} \frac{y_{n}+\tau f_{2}(x_{n}, y_{n})}{y_{n}+\tau g_{2}(x_{n}, y_{n})},
    \label{eqn:0-1-1a}
\end{equation}
where 
\begin{equation}
    \begin{aligned}
        f_1(x,y)=y, \;\;
        g_1(x,y)=x, \;\; \\
        f_2(x,y)=\frac{1}{1+x^m}, \;\;
        g_2(x,y)=\frac{y}{b}.
    \end{aligned}
    \label{eqn:elements1}
\end{equation} 
It is easily found that eq.(\ref{eqn:0-1-1a}) becomes 
$\frac{dx}{dt}=f_1(x,y)-g_1(x,y)$,
$\frac{dy}{dt}=f_2(x,y)-g_2(x,y)$ in the limit of $\tau \to 0$.
Equation (\ref{eqn:00a}) has a positive fixed point $(\bar x,\bar y)$, 
where $\bar x=\bar y =\frac{b}{(1+\bar y^m)}>0$ holds.
Note that $(\bar x,\bar y)$ also becomes the fixed point of eq. (\ref{eqn:trop1}) for arbitrary $\tau$.
The Jacobi matrix for eq. (\ref{eqn:00a}) at $(\bar x,\bar y)$ is given by 
$
J = \bigl(
\begin{smallmatrix}
   -1 & 1 \\
   \bar z & -b^{-1}
\end{smallmatrix}
\bigl)
$
where $\displaystyle \bar z \equiv \frac{\partial f_2}{\partial x}(\bar x,\bar y)=-\frac{m\bar y ^{m+1}}{b^2}$.
Note that $\bar z<0$.
The trace $T$ and determinant $\Delta$ of $J$ are 
$T\equiv$ Tr $J =-(b^{-1}+1)<0$ and 
$\Delta \equiv$ det $J=b^{-1}-\bar z>0$, respectively.

Here we consider the function $P_{nd}(\tau)$ given as
\begin{equation}
  P_{nd}(\tau) \equiv A_{nd} \tau^2  + B_{nd} \tau + C_{nd},
  \label{eqn:spiral_ineq}  
\end{equation}
where 
\begin{equation}
    \label{eqn:spiral_ineq_1}
    \begin{aligned}
        A_{nd} & = F^2
        -4\bar{x}\bar{y}\bar{f}_{1}\bar{f}_{2}\Delta, \\
        B_{nd} & = 2\bar{x}\bar{y} T F 
        -4\bar{x} \bar{y} \left( \bar{x} \bar{f}_{2} + \bar{y} \bar{f}_{1} \right) \Delta, \\
        C_{nd} & = \left( \bar{x} \bar{y} \right)^2
        \left( T^2 - 4\Delta \right), 
    \end{aligned}
\end{equation}
$F \equiv \left(\bar{x} \bar{f}_{2} J_{11}+\bar{y} \bar{f}_{1} J_{22} \right)$, and $J_{ij}$ denotes the $(i, j)$ component 
of the matrix $J$ $(i,j=1,2)$\cite{Ohmori2023}.
The sign of $P_{nd}(\tau)$ determines 
whether the fixed point $(\bar x,\bar y)$ is spiral or not.
Now $A_{nd} = 4\bar x ^4b^{-1}\bar z < 0$ and
$B_{nd} = 4\bar x ^4\bar z(1+b^{-1}) < 0$.
If $C_{nd} < 0$, where the fixed point $(\bar x, \bar y)$ is spiral in eq.(\ref{eqn:00a}), $T^2<4 \Delta$, 
then $P_{nd}(\tau) < 0$ holds for all $\tau$ 
and the fixed point $(\bar x,\bar y)$ is also spiral in eq. (\ref{eqn:trop1}) for all $\tau >0$.
On the other hand, when $C_{nd} > 0$, 
$(\bar x,\bar y)$ becomes a spiral for $\tau$ 
satisfying $P_{nd}(\tau) < 0$.

Regarding the stability of $(\bar x,\bar y)$, 
the following values of $\alpha$ and $\beta$ are focused on\cite{Ohmori2023}:
\begin{equation}
    \label{eqn:stable_ineq_a}
    \begin{aligned}
        \alpha & =  \Delta \bar{x} \bar{y} + F
        = - \bar{x}^2 \left( b^{-1}+\bar z \right), \\ 
        \beta & =  T\bar{x} \bar{y} = -\bar{x}^2 \left( b^{-1}+1 \right) < 0.
    \end{aligned}
\end{equation}
Note that the sign of $\alpha$ 
determines the stability of the spiral fixed point.
When $\alpha<0$, or $b^{-1}+\bar z>0$, $(\bar x,\bar y)$ is stable for any $\tau>0$.
If $0<m\leq 1$, then $b^{-1}+\bar z>0$ is always satisfied for any $\tau>0$, 
and $(\bar x,\bar y)$ becomes stable.
On the other hand for $m>1$, when $b^{-1}+\bar z<0$, 
$(\bar x,\bar y)$ is stable (unstable) for $0<\tau<\gamma$ 
($\tau>\gamma$), respectively, 
where $\displaystyle\gamma \equiv -\frac{\beta}{\alpha}=-\frac{b^{-1}+1}{b^{-1}+\bar z}$.
Therefore, at $\tau = \gamma$, the Neimark-Sacker bifurcation occurs.
Note that these results for the spiral fixed points are consistent with the previous studies done by Gibo and Ito\cite{Gibo2015}.

\begin{figure}[b!]
    \begin{center}
        \includegraphics[height=2.7cm]{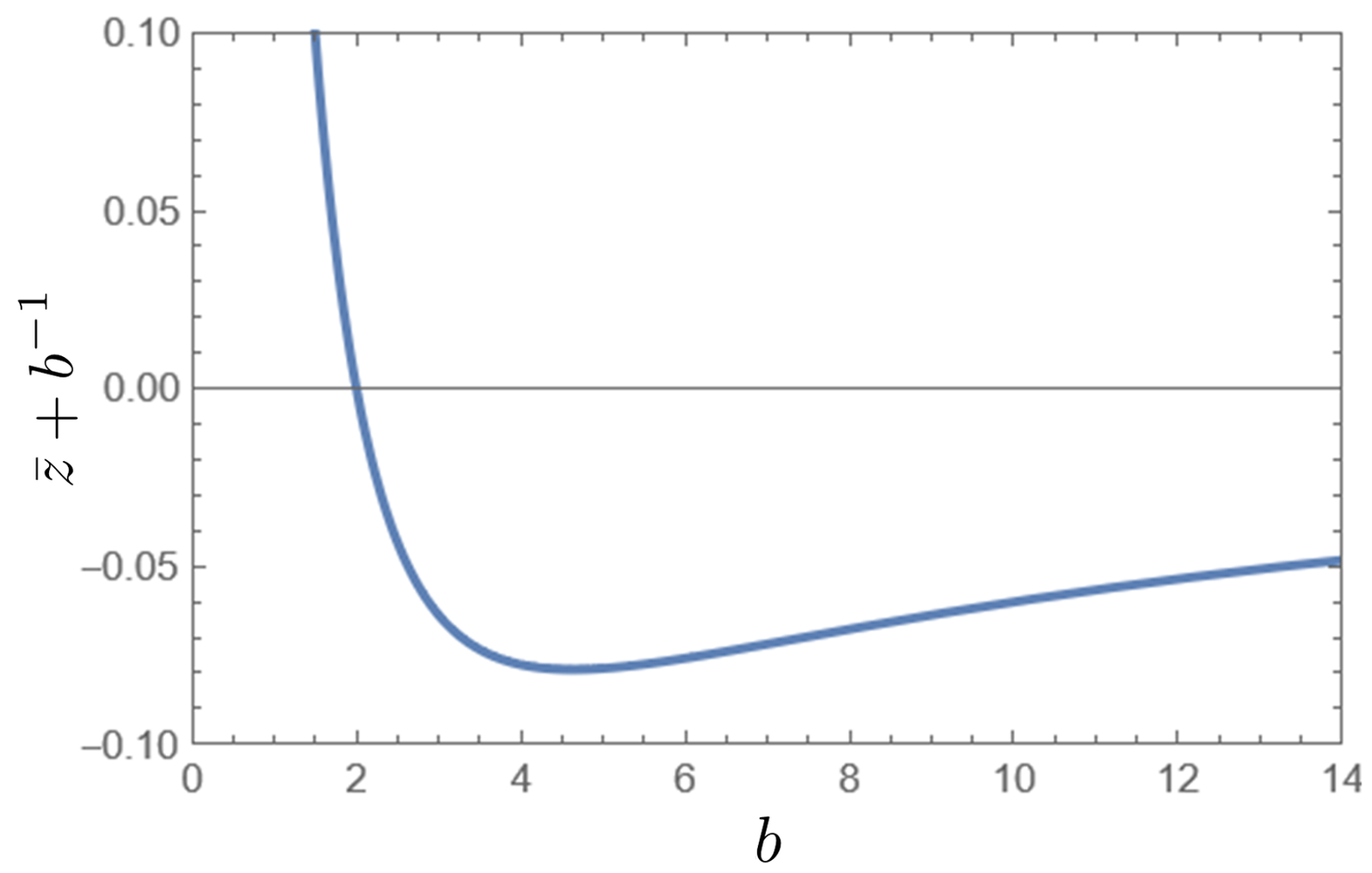}
        \includegraphics[height=2.7cm]{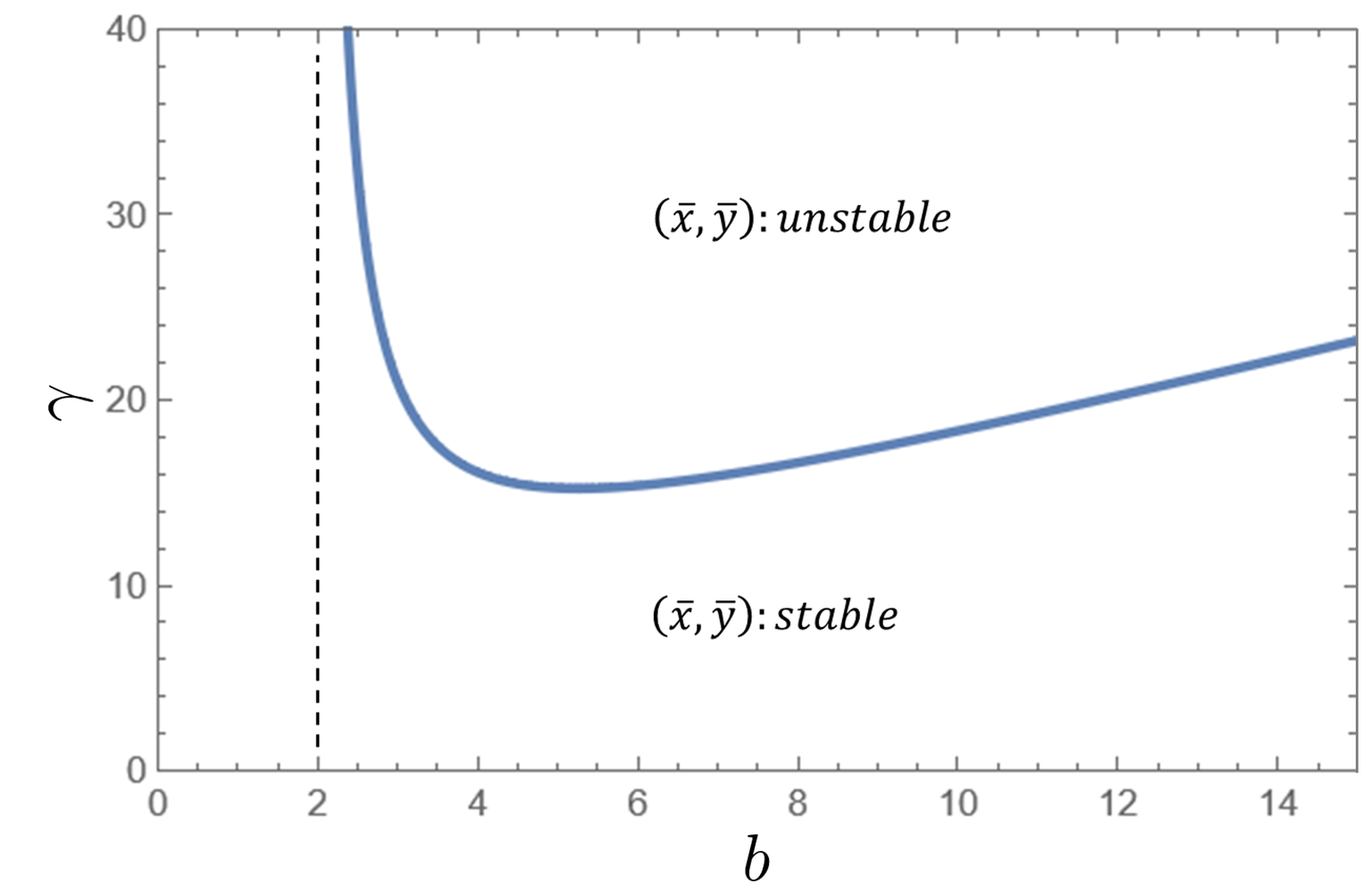}
        \\
        \hspace{0.6cm}
        (a)
        \hspace{3.2cm}
        (b)\\
        \caption{\label{Fig.Bif_Curves} 
            The graphs of (a) $b^{-1}+\bar z(b)$ and (b) $\gamma (b) =-\frac{b^{-1}+1}{b^{-1}+\bar{z}(b)}$. We set $m=2$.
        }
    \end{center}
\end{figure}

\begin{figure}[b!]
    \begin{center}
        \includegraphics[width=4.2cm]{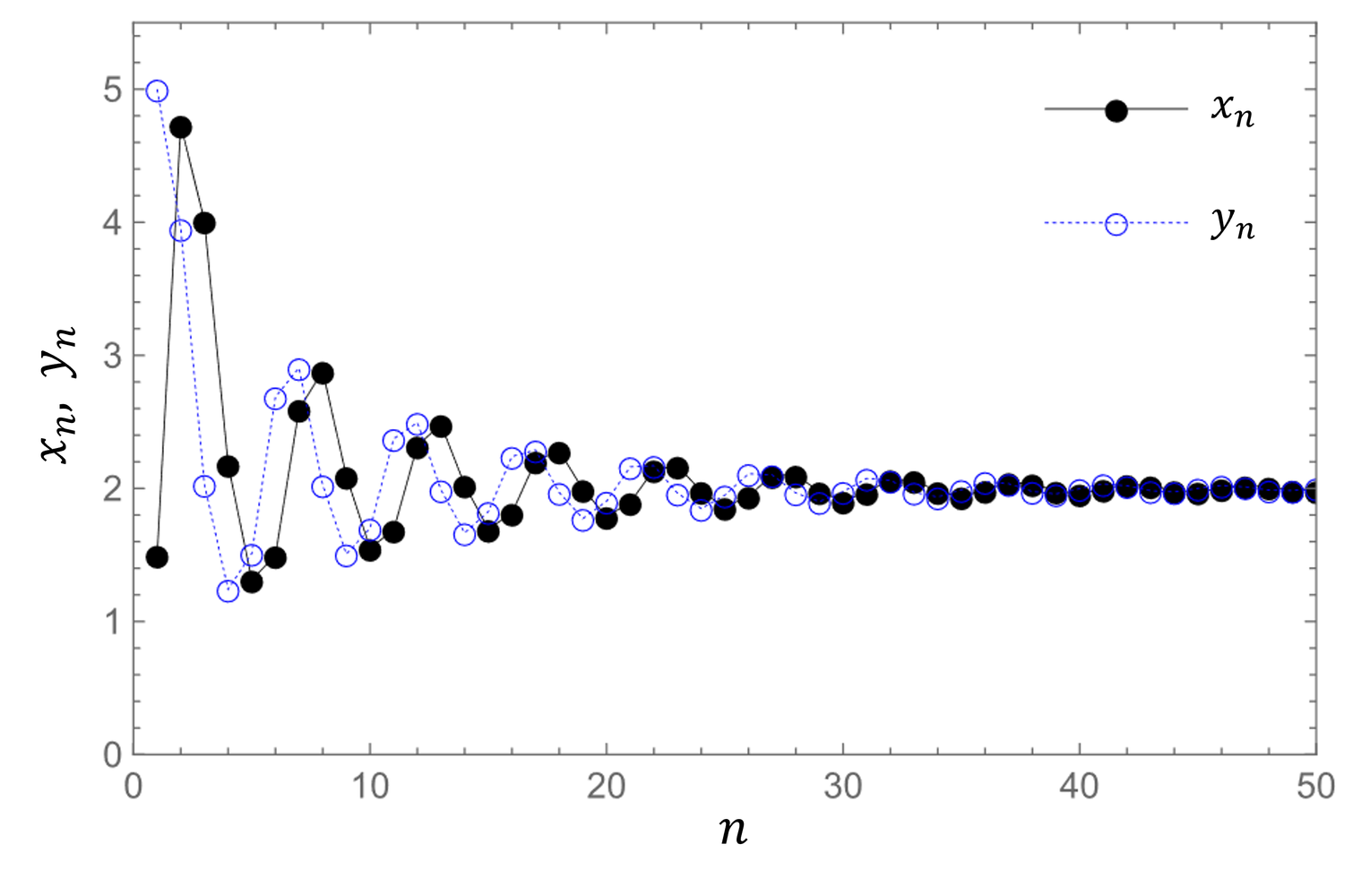}
        \includegraphics[width=4.2cm]{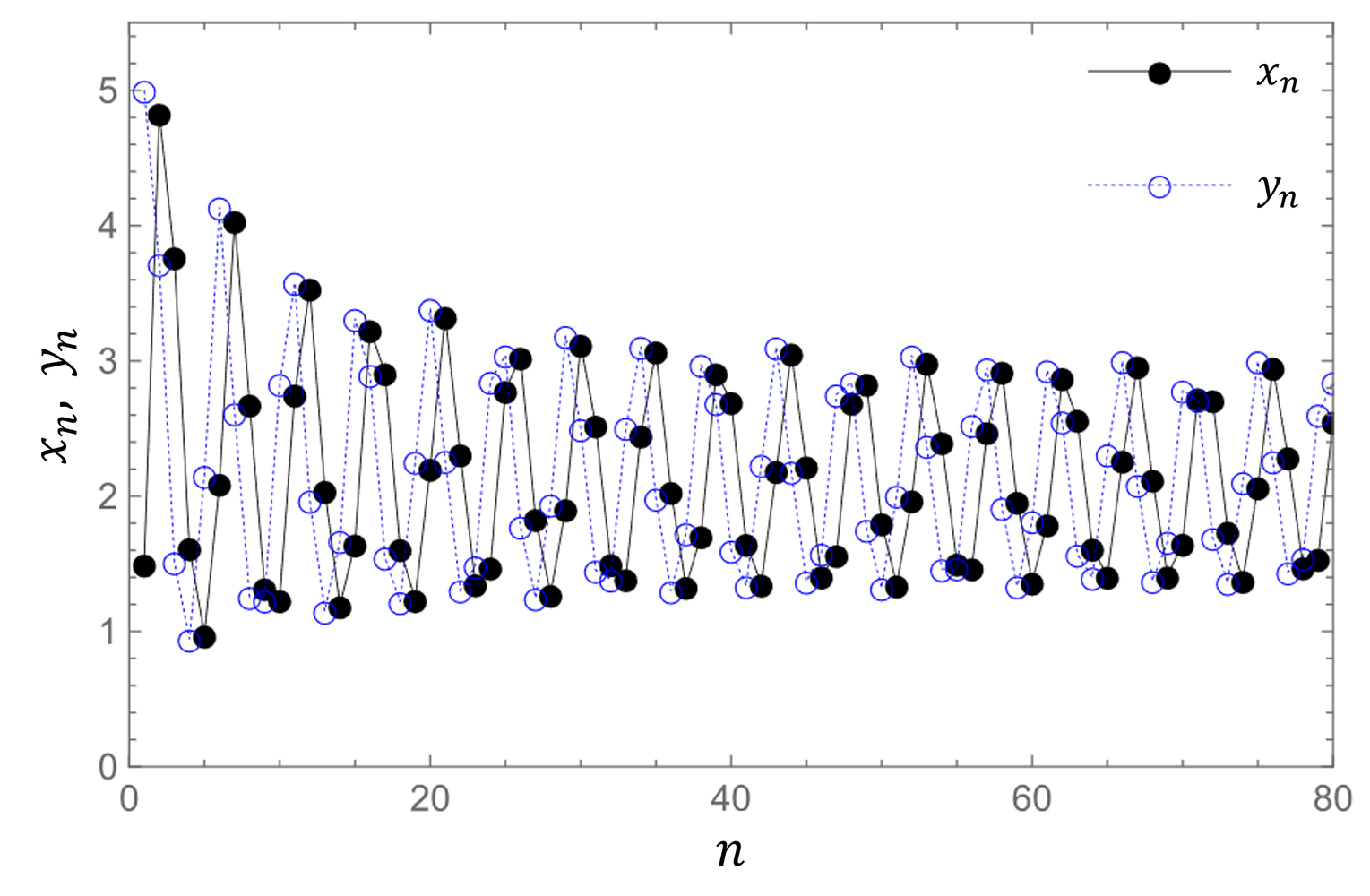}
        \\
        \hspace{0.3cm}
        (a)
        \hspace{3.5cm}
        (b)\\
        \caption{\label{Fig.TDE_time_evol.} 
            Time evolutions of eq. (\ref{eqn:trop1}) for $m=2, b=10$. (a) $\tau =12$, (b) $\tau =20$.
        }
    \end{center}
\end{figure}

Now we set $m=2$ and $b=10$ as an example.
In this example, we obtain $(\bar x,\bar y)=(2,2)$, 
$A_{nd}=-1.024$, $B_{nd}=-11.264$, and $C_{nd}=+2.72$.
Then $(\bar x,\bar y)$ becomes spiral 
for $\tau > 0.236397\cdots$.
Figures \ref{Fig.Bif_Curves} (a) and (b) show the graphs of $b^{-1}+\bar z(b)$ and $\gamma (b) =-\frac{b^{-1}+1}{b^{-1}+\bar{z}(b)}$, respectively.
From Fig.\ref{Fig.Bif_Curves}(a), it is found that $b^{-1}+\bar z(b)<0$ holds when $b>2$. 
The Neimark-Sacker bifurcation occurs at $b = 2$, 
and the limit cycle solutions can emerge in the region $\tau>\gamma (b)$ as shown in Fig.\ref{Fig.Bif_Curves}(b).
Figure \ref{Fig.TDE_time_evol.} shows the time evolution of eq. (\ref{eqn:trop1}) 
from the initial state $(x_0,y_0)=(1.5,5)$ when (a) $\tau =12$ and (b) $\tau =20$.
Since $\gamma = \frac{55}{3}=18.333\dots$ for $m=2$ and $b=10$, 
$(x_{n}, y_{n})$ converges to $(\bar x,\bar y)$ 
for $\tau=12$ as shown in Fig.\ref{Fig.TDE_time_evol.}(a).
On the other hand, a cyclic solution is obtained 
around $(\bar x,\bar y)$ for $\tau=20$ 
as shown in Fig.\ref{Fig.TDE_time_evol.}(b).  
\begin{figure}[b!]
    \begin{center}
        \includegraphics[width=2.7cm]{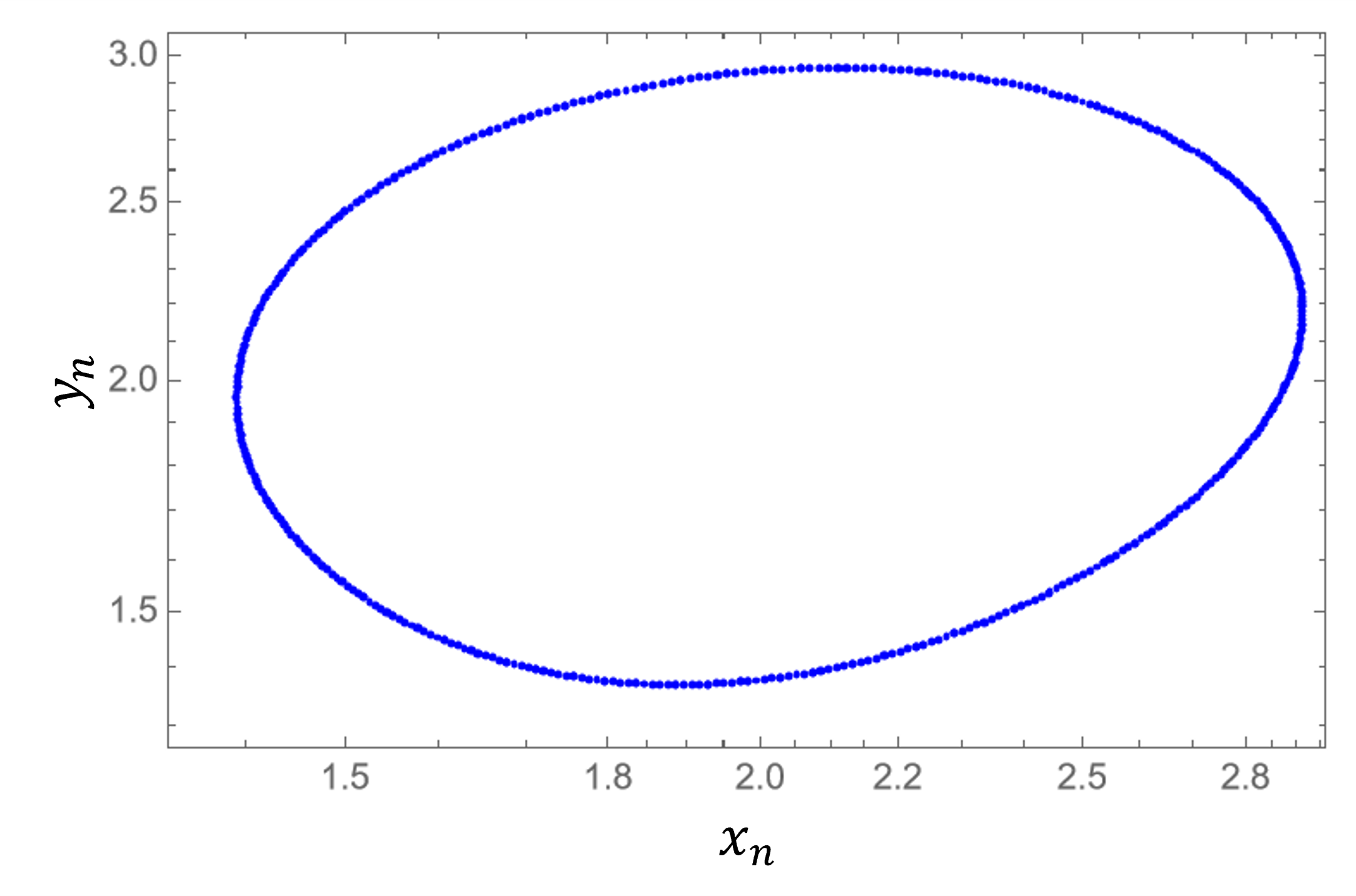}
        \includegraphics[width=2.7cm]{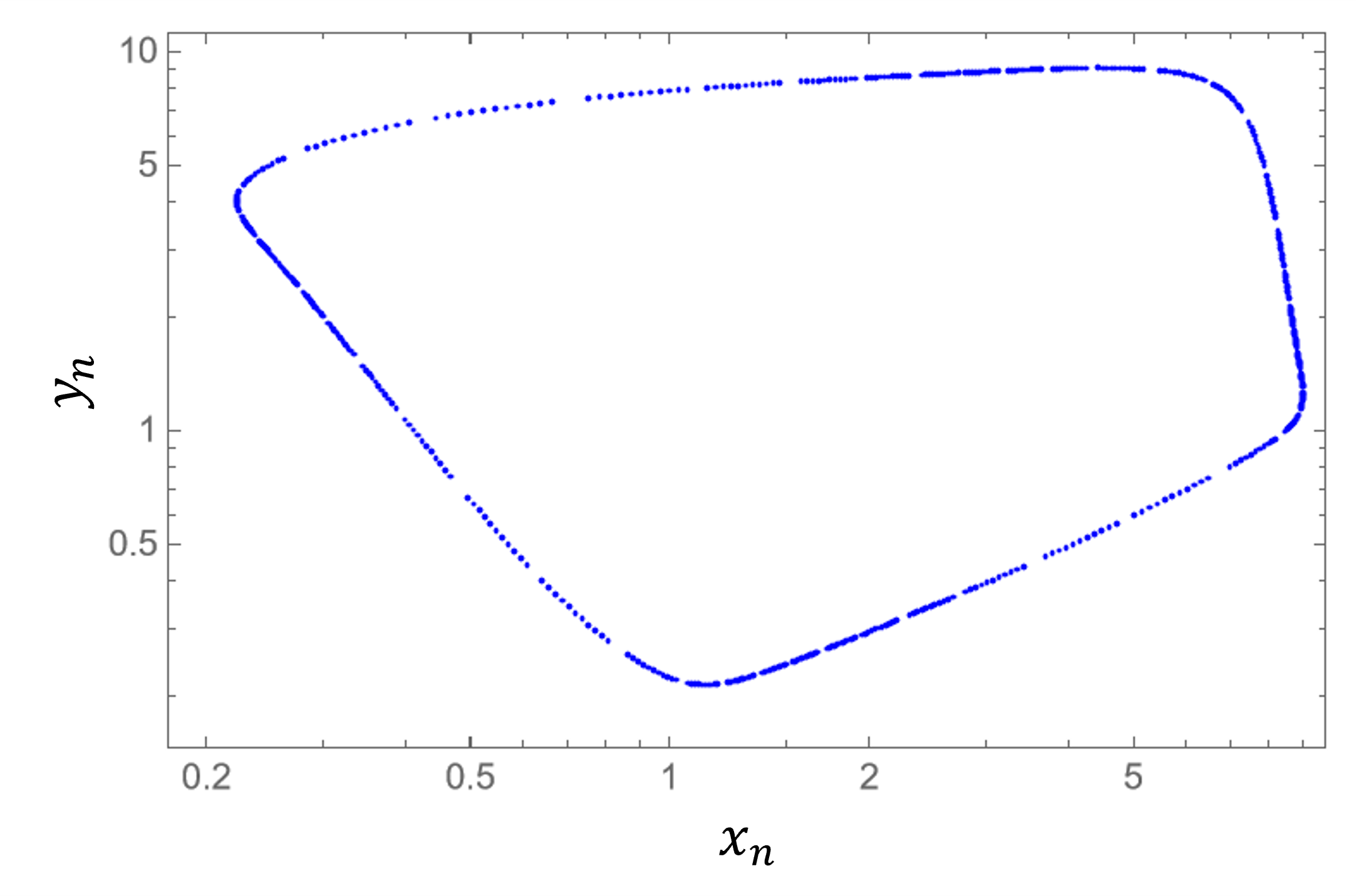}
        \includegraphics[width=2.7cm]{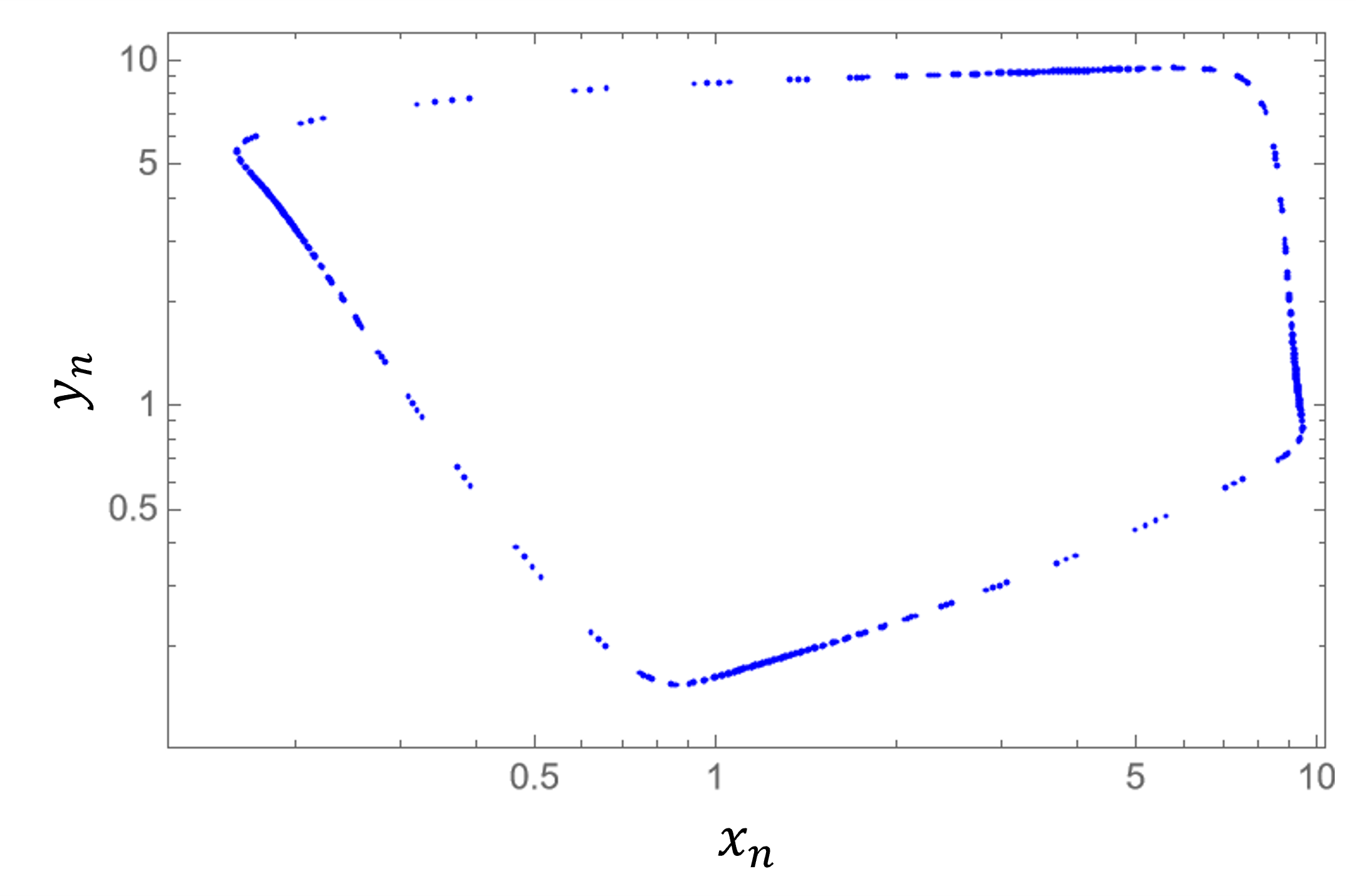}
        \\
        \hspace{0.2cm}
        (a) $\tau = 20$
        \hspace{1cm}
        (b) $\tau = 100$
        \hspace{0.8cm}
        (c) $\tau = 150$\\
        \includegraphics[width=2.7cm]{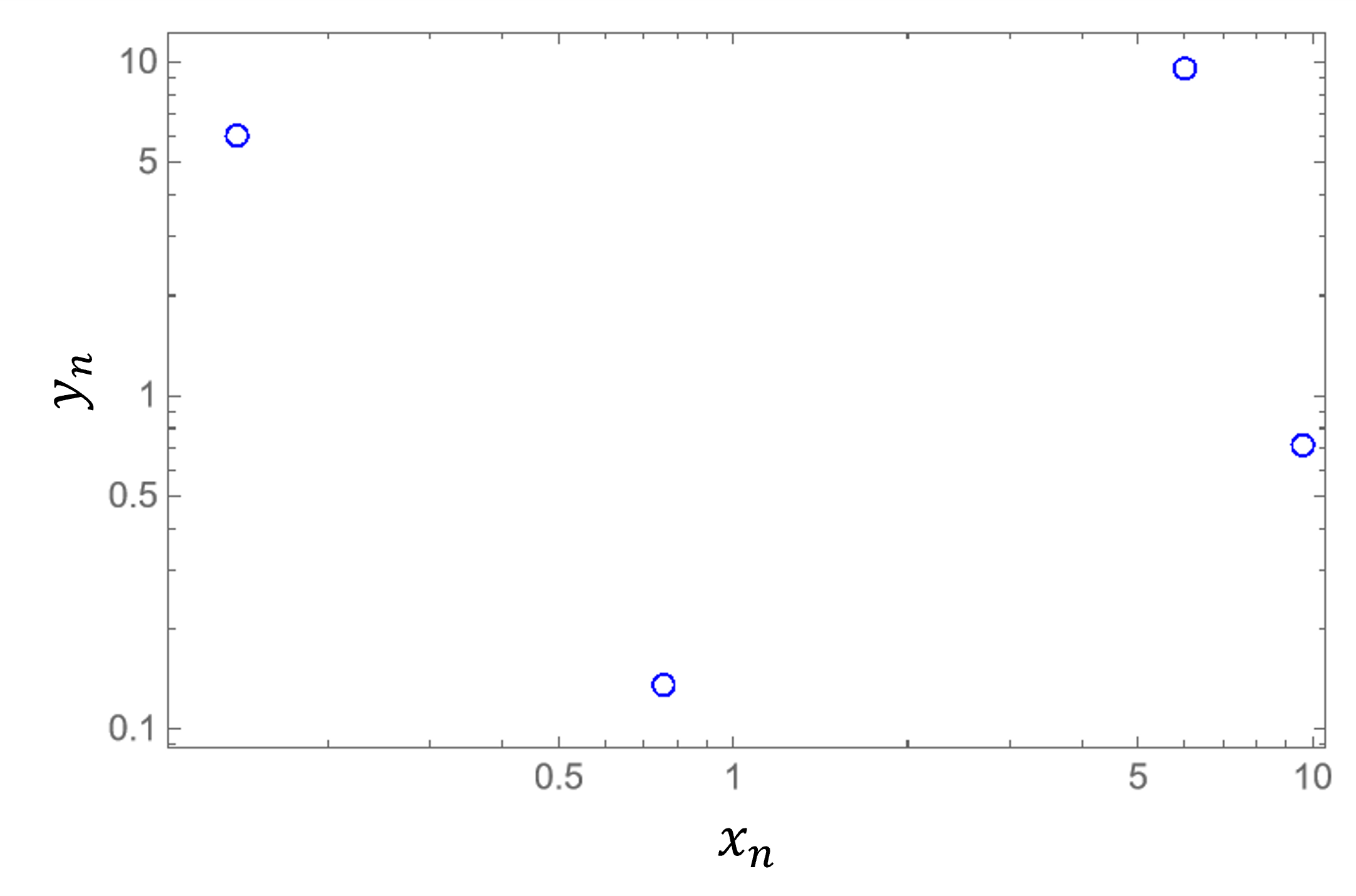}
        \includegraphics[width=2.7cm]{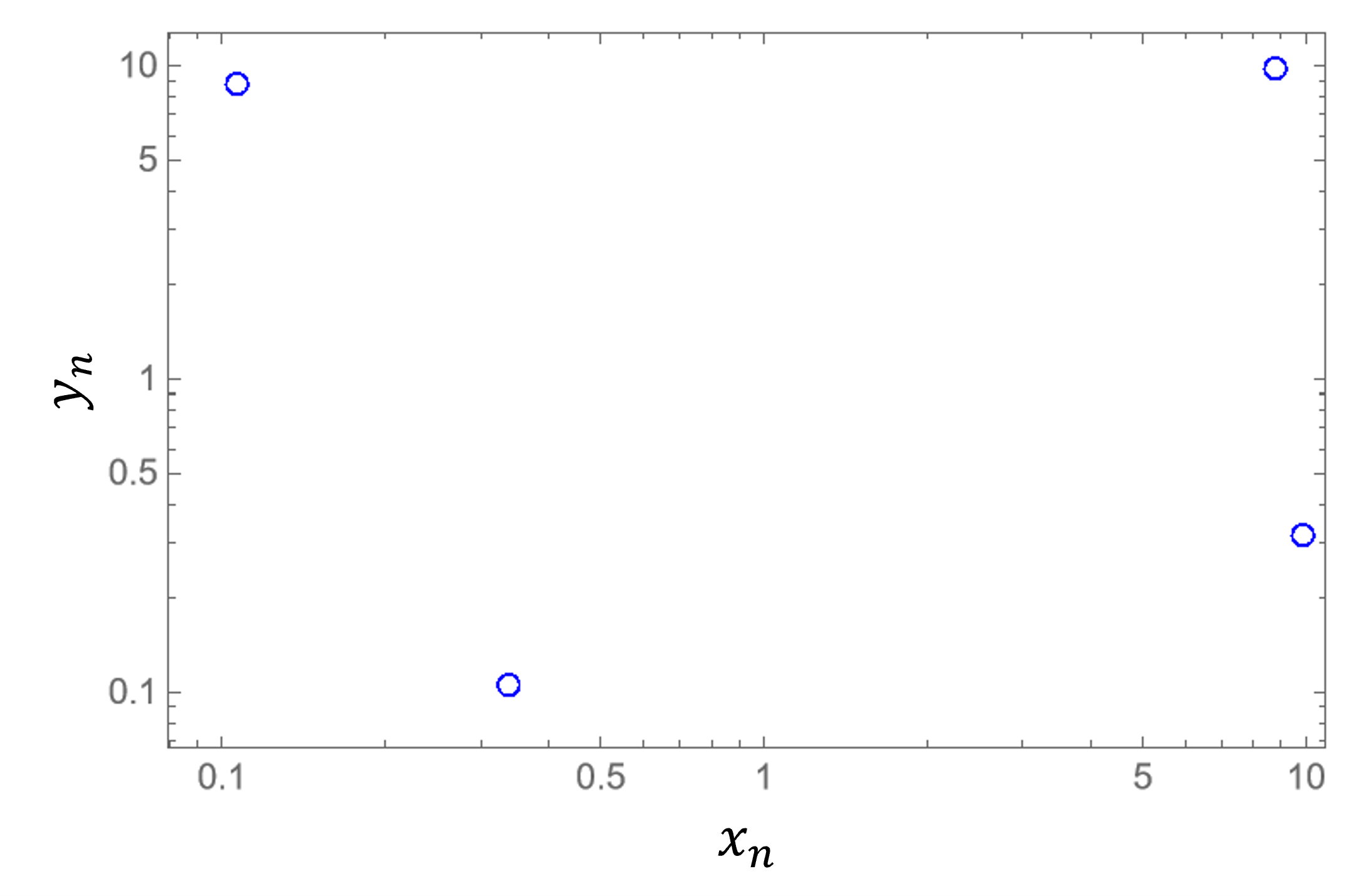}
        \includegraphics[width=2.7cm]{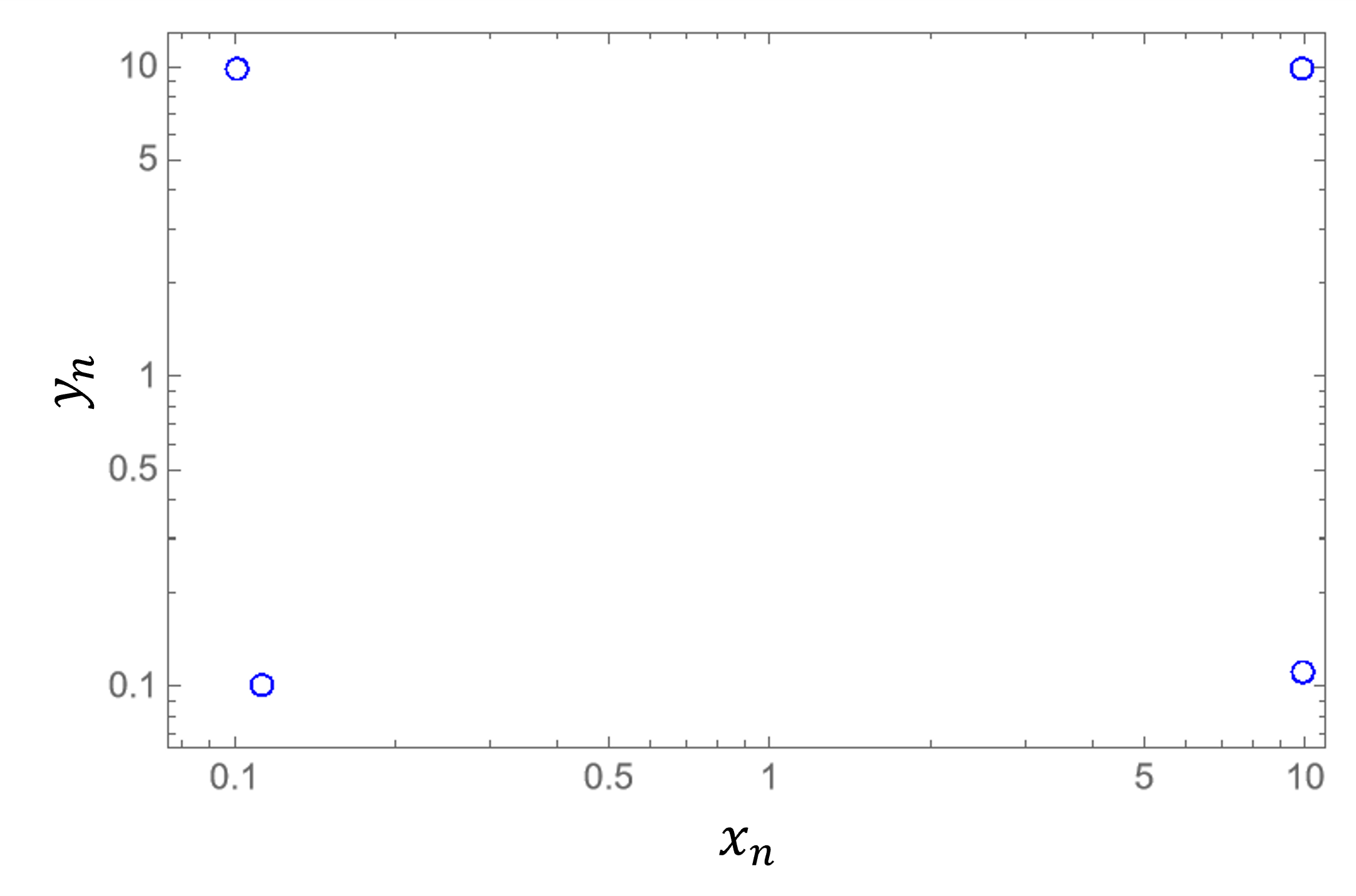}
        \\
        \hspace{0.1cm}
        (d) $\tau = 200$
        \hspace{0.8cm}
        (e) $\tau = 500$
        \hspace{0.8cm}
        (f) $\tau = 10^4$\\
        \caption{\label{Fig.Transition}
            $\tau$-dependence of the states $(x_{n}, y_{n})$
            in the limit cycles obtained from eq. (\ref{eqn:trop1}) 
            for $m=2$ and $b=10$.
        }
    \end{center}
\end{figure}
%

For the states in the limit cycle solutions, 
Fig.\ref{Fig.Transition} shows the plot of $(x_{n}, y_{n})$ 
as a function of $\tau$.
From this figure, we confirm the following features.
(i) When $\tau = 20$, the states $(x_{n}, y_{n})$ are broadly distributed 
as shown in Fig.\ref{Fig.Transition}(a).
(ii) From Fig.\ref{Fig.Transition}(b)-(c), as $\tau$ increases 
the states tend to become sparsely distributed. 
(iii) When $\tau$ is larger than 200,  
only four states exist in the limit cycles 
as shown in Fig.\ref{Fig.Transition}(d)-(f).
We can consider the limit cycles with the four discrete states 
as the ultradiscrete limit cycles.
Then our result shows that the ultradiscrete limit cycle 
emerges at a finite value of $\tau$.
%
%
Also it is found that the ultradiscrete limit cycle 
with only four states is realized for large $\tau$ 
even in the case of $\tau \to \infty$. 
Similar emergence of ultradiscrete limit cycles 
for a large vaule of $\tau$ has already been found in the case of the ultradiscrete Sel'kov model\cite{Ohmori2021}.
%
%

\section{Max-plus modelling}

Here we discuss dynamical properties 
of the ultradiscrete limit cycle when $\tau$ is infinity.
In the case of $\tau \to \infty$, eq.(\ref{eqn:trop1}) becomes
\begin{equation}
    x_{n+1} = y_n, 
    \;\;\;\;
    y_{n+1} = \frac{b}{1+x_n^m}.
    \label{eqn:trop1_inf} 
\end{equation}
Perfoming the variable transformations,
$x_n=e^{X_n/\varepsilon}$, $y_n=e^{Y_n/\varepsilon}$, $b=e^{B/\varepsilon}$, 
and taking the ultradiscrete limit\cite{Tokihiro2004}
\[
    \displaystyle\lim_{\varepsilon  \to +0} \varepsilon  \log(e^{A/\varepsilon }+e^{B/\varepsilon }+\cdot \cdot \cdot )
    =\max(A,B,\cdot \cdot \cdot), 	
\]
we obtain the max-plus equation, %
\begin{equation}
    X_{n+1} = Y_n, 
    \;\;\;\;
    Y_{n+1} = B-\max(0,m X_n).
    \label{eqn:1-1a} 
\end{equation} 
Gibo and Ito have derived essentially the same equation  
as eq.(\ref{eqn:1-1a}) and reported existence of the ultradiscrete limit cycle\cite{Gibo2015}. 
Here we report results of further investigation for the dynamical properties of eq.(\ref{eqn:1-1a}).
%

%

%
%

%
%

%
%
%
%

%


%
When $X_n > 0$, eq.(\ref{eqn:1-1a}) can be rewritten as  
\begin{equation}
    \begin{pmatrix}
        X_{n+1}  \\
        Y_{n+1}  
    \end{pmatrix}
    = 
    \begin{pmatrix}
        0 & 1  \\
        -m & 0 
    \end{pmatrix}
    \begin{pmatrix}
        X_{n}  \\
        Y_{n} 
    \end{pmatrix}
    +
    \begin{pmatrix}
        0  \\
        B  
    \end{pmatrix}, 
    \label{eqn:2-2a}
\end{equation} 
and eq.(\ref{eqn:2-2a}) has the fixed point $\bm{\bar x}_{\rm I}=\left(\frac{B}{1+m},\frac{B}{1+m}\right)$.
The trace and the determinant of the matrix
$
    \bm{A}_{\rm I}=\left(
    \begin{smallmatrix}
        0 & 1  \\
        -m & 0  \\
    \end{smallmatrix}
    \right)
$
are given as Tr$\bm{A}_{\rm I} = 0$ and det$\bm{A}_{\rm I} = m$, respectively.
Therefore, the discrete trajectory given by eq.(\ref{eqn:2-2a}) is characterized as spiral sink $(0<m<1)$, center $(m=1)$, and spiral source $(1<m)$, respectively\cite{Galor}. 
(In all cases, rotations are in the clockwise direction.)
When $X_n < 0$, eq.(\ref{eqn:1-1a}) has the matrix form  
\begin{eqnarray}
    \begin{pmatrix}
        X_{n+1}  \\
        Y_{n+1}  
    \end{pmatrix}
    = 
    \begin{pmatrix}
        0 & 1  \\
        0 & 0  
    \end{pmatrix}
    \begin{pmatrix}
        X_{n}  \\
        Y_{n}  
    \end{pmatrix}
    +
    \begin{pmatrix}
        0  \\
        B  
    \end{pmatrix}.
    \label{eqn:2-2b}
\end{eqnarray} 
Equation (\ref{eqn:2-2b}) has the fixed point $\bm{\bar x}_{\rm II}=(B,B)$, which is a stable node.
Then the dynamics of eq.(\ref{eqn:1-1a}) can be characterized by eqs.(\ref{eqn:2-2a}) and (\ref{eqn:2-2b}).

For  eqs.(\ref{eqn:2-2a})-(\ref{eqn:2-2b}), 
we first consider $B < 0$.
The time evolution of $(X_n,Y_n) \equiv \bm{x}_{n}$ 
from the initial condition $(X_0,Y_0) \equiv \bm{x}_{0}$
can be summarized dependent on the signs of $X_0$ and $Y_0$
as shown in Table \ref{Tbl.1}.
Thus, $\bm{\bar x}_{\rm II}=(B,B)$ is stable and every initial state converges to $\bm{\bar x}_{\rm II}$ at most four iteration steps, for any $m>0$.
%
%
%

%

\begin{figure}[t!]
    \begin{center}
        \includegraphics[width=4.2cm]{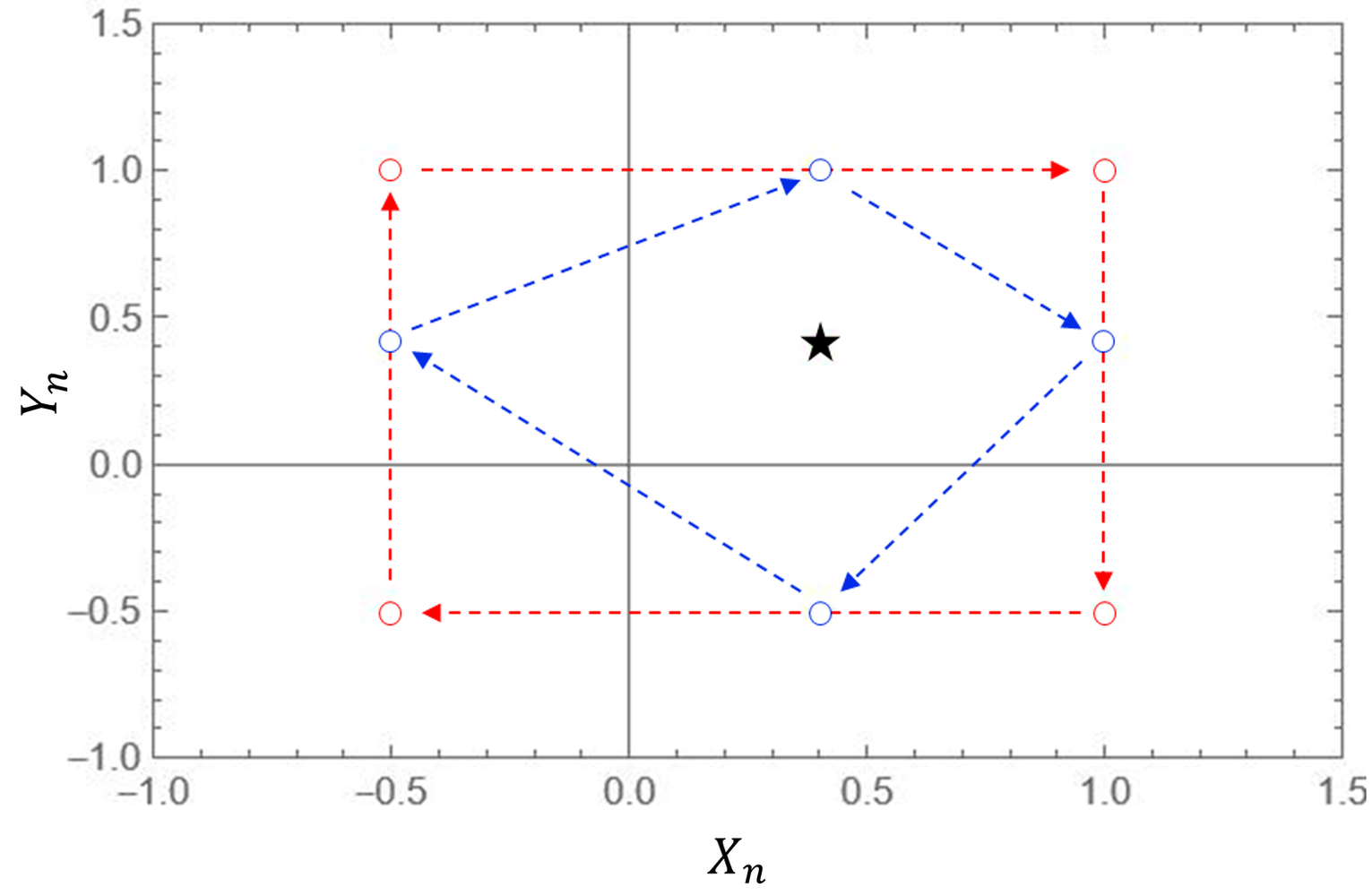}
        \includegraphics[width=4.2cm]{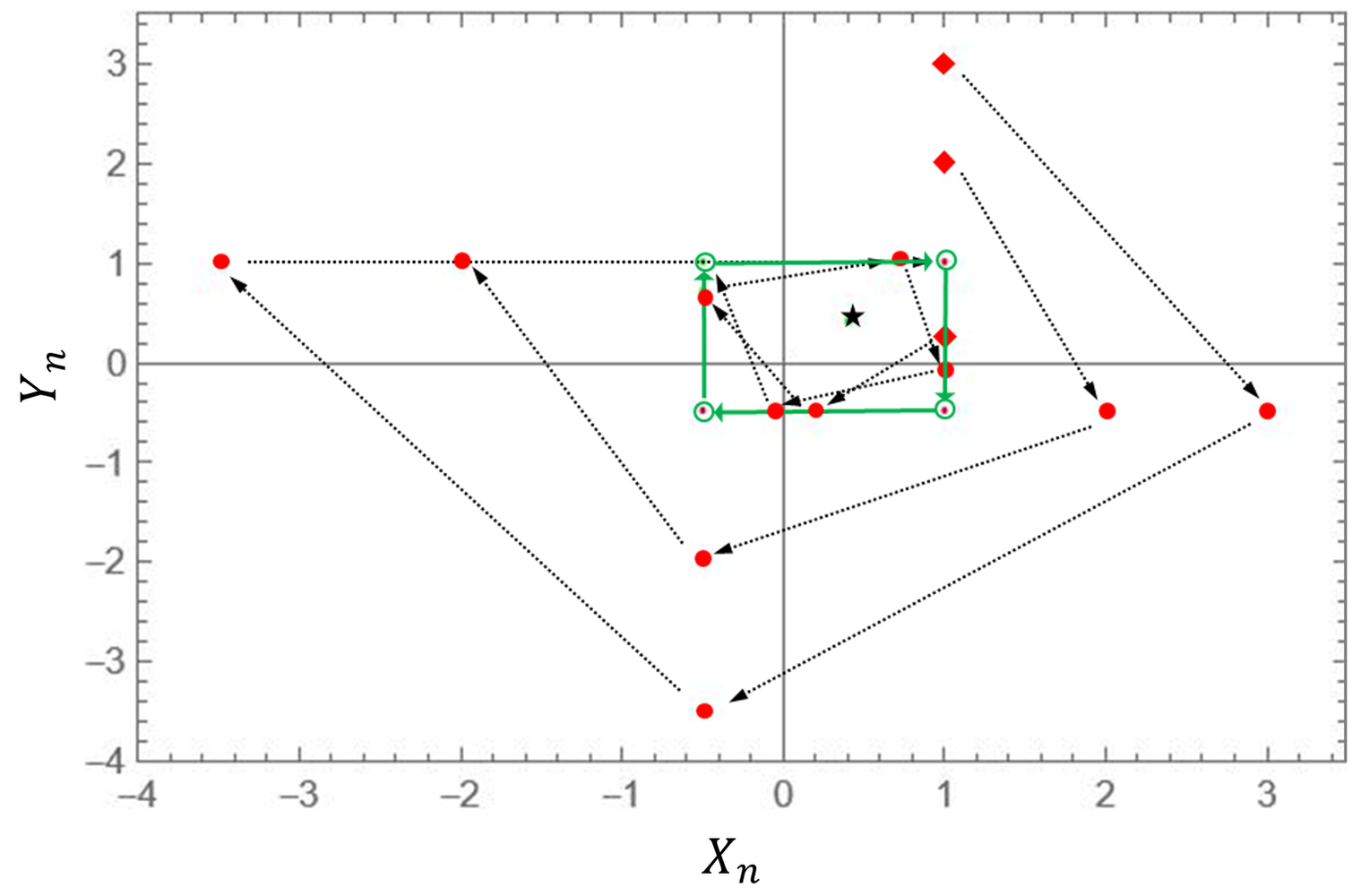}
        \\
        \hspace{0.3cm}
        (a)
        \hspace{3.5cm}
        (b)\\
        \caption{\label{fig:lc} 
            (a) The two limit cycles $\mathcal{C}$ (red circles) 
            and $\mathcal{C}_s$ (blue circles).
            We set $B=1$ and $m=1.5$.
            (b) Examples of trajectories starting 
            from three different filled squares.
            The trajectories finally converge into $\mathcal{C}$ 
            consisting of the four green open circles.
            The star in each figure stands 
            for the fixed point 
            $\bm{\bar x}_{I}=\left(\frac{B}{1+m},\frac{B}{1+m}\right)$. 
        }
    \end{center}
\end{figure}

Next we set $B>0$, where 
$\bm{\bar x}_{\rm{I}}=\left(\frac{B}{1+m},\frac{B}{1+m}\right)$ is a unique unstable fixed point.
When $m > 1$, it is found that there exist the two clockwise periodic solutions around $\bm{\bar x}_{\rm I}$, 
$\mathcal{C}$ and $\mathcal{C}_s$, as shown in Fig. \ref{fig:lc} (a); they are composed of the following four points: 
\begin{align*}
    \mathcal{C} : & (B,B) [\equiv \bm{x}_{0}^{\mathcal{C}}] 
        \rightarrow (B,(1-m)B)  \\
        & \rightarrow ((1-m)B,(1-m)B) 
        \rightarrow ((1-m)B,B) 
        [ \rightarrow \bm{x}_{0}^{\mathcal{C}} ],\\
    \mathcal{C}_s : & \left(B/(m+1),B\right) [\equiv \bm{x}_{0}^{\mathcal{C}_s}]
        \rightarrow \left(B, B/(m+1)\right) \\
        & \rightarrow \left(B/(m+1),(1-m)B \right) \\
        & \rightarrow \left((1-m)B, B/(m+1) \right) 
        [ \rightarrow \bm{x}_{0}^{\mathcal{C}_s} ] .
\end{align*}
Figure \ref{fig:lc} (b) shows trajectories 
from three different initial conditions; 
they all finally converge to $\mathcal{C}$.
\begin{table}[t!]
    \caption{The time evolution of $\bm{x}_{n}=(X_n,Y_n)$ 
        from the initial condition $\bm{x}_{0}=(X_0,Y_0)$ for $B<0$.
        The signs ``$+$'' and ``$-$'' in the table represent the signs of $X_n$ and $Y_n$. 
    }
    \begin{center}
        \begin{tabular}{|c|c|c|}
        \hline
          & (a) & (b) \\
         $n$ & $X_{n}~~Y_{n}$ & $X_{n}~~Y_{n}$ \\
        \hline
        $0$ & $+~~~+$ & $-~~~+$ \\
        %
        $1$ & $+~~~-$ & $+~~~B$ \\
        %
        $2$ & $-~~~-$ & $B~~~-$ \\
        %
        $3$ & $-~~~B$ & $-~~~B$ \\
        %
        $4$ & $B~~~B$ & $B~~~B$ \\
        \hline
        \end{tabular}
    \end{center}
    \label{Tbl.1}
\end{table}
%
%
%
%
%

%

\begin{figure}[t!]
    \begin{center}
        \includegraphics[width=4.2cm]{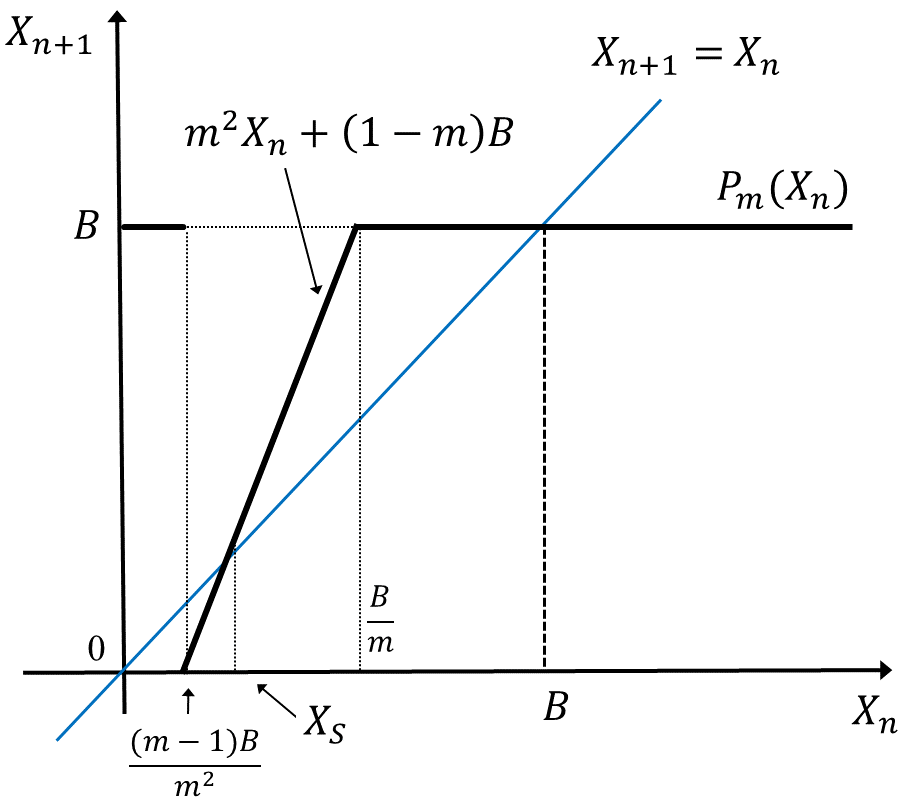}
        \includegraphics[width=4.2cm]{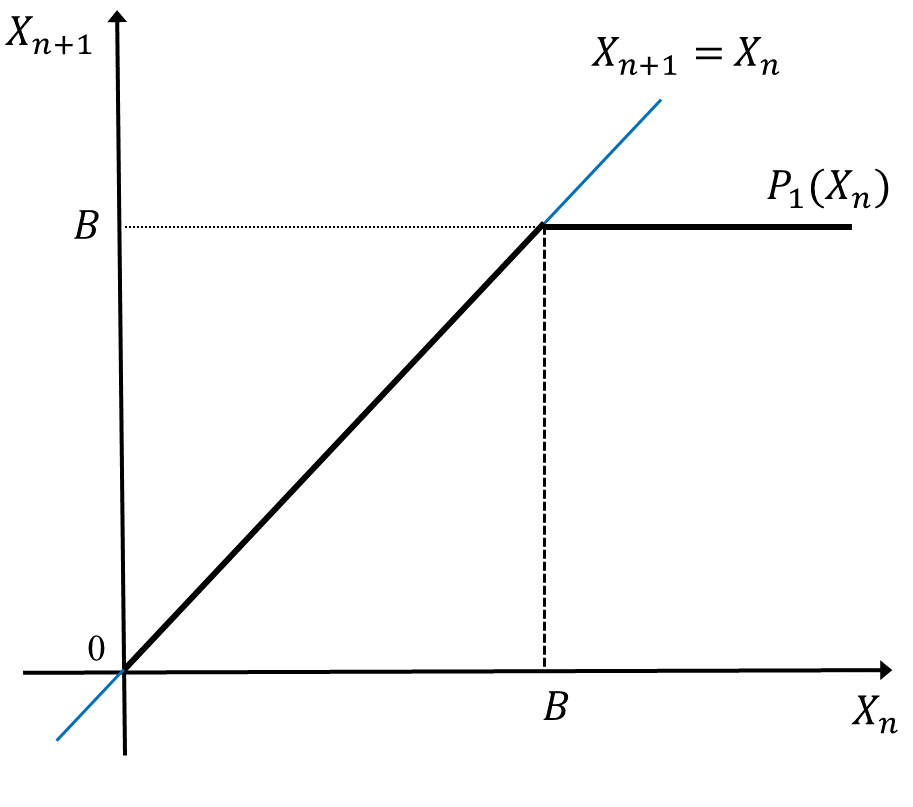}
        \\
        \hspace{0.3cm}
        (a)
        \hspace{3.5cm}
        (b)\\
        \caption{\label{Fig.P_map} 
            The graphs of the Poincar\'{e} map $X_{n+1}=P_m(X_{n})$, eq.(\ref{eqn:Pmap}), 
            for (a) $m>1$ and (b) $m=1$.
            The blue line shows $X_{n+1}=X_n$.
        }
    \end{center}
\end{figure}

\begin{figure}[t!]
    \begin{center}
        \includegraphics[width=4.2cm]{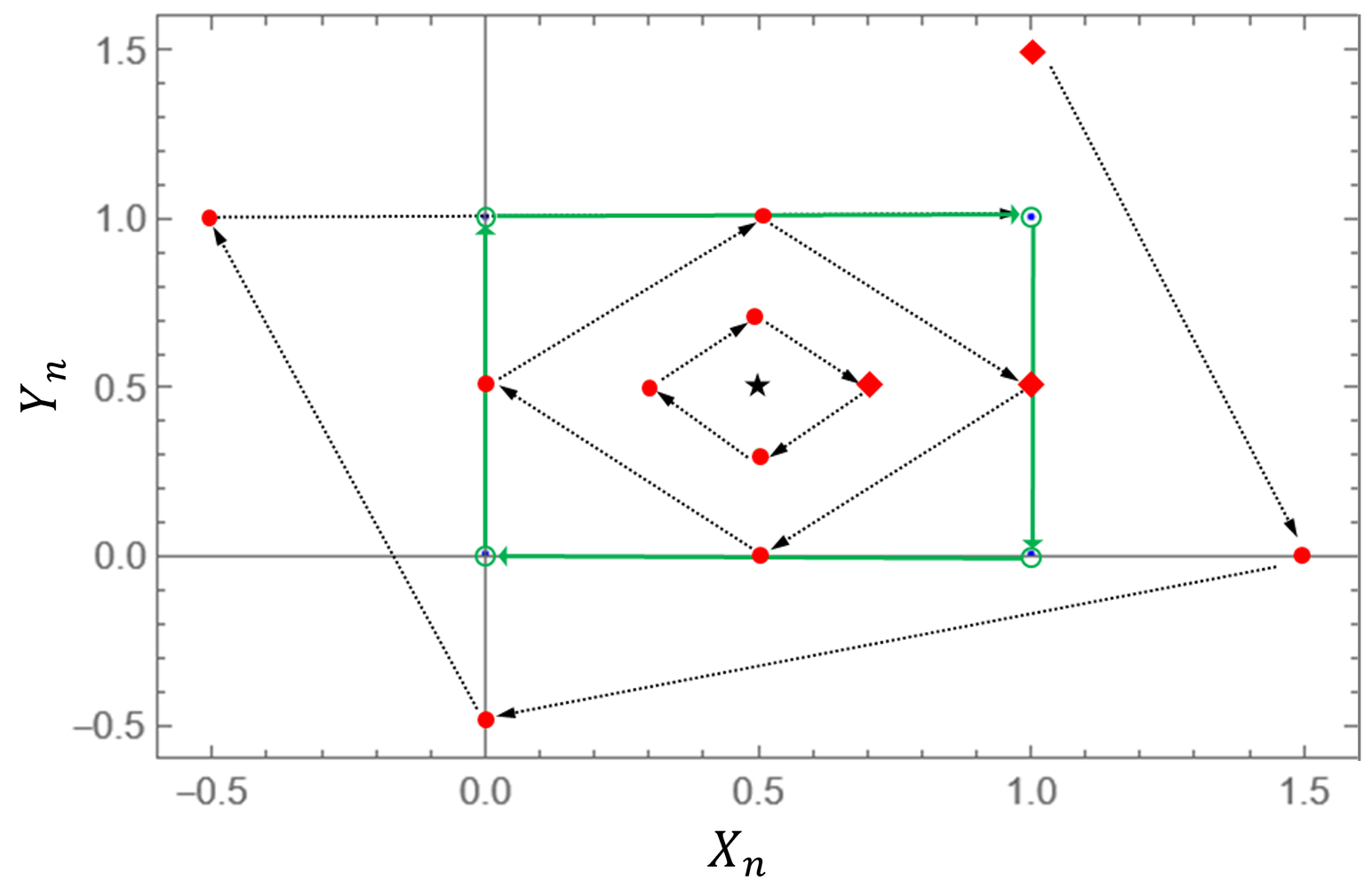}
        \includegraphics[width=4.2cm]{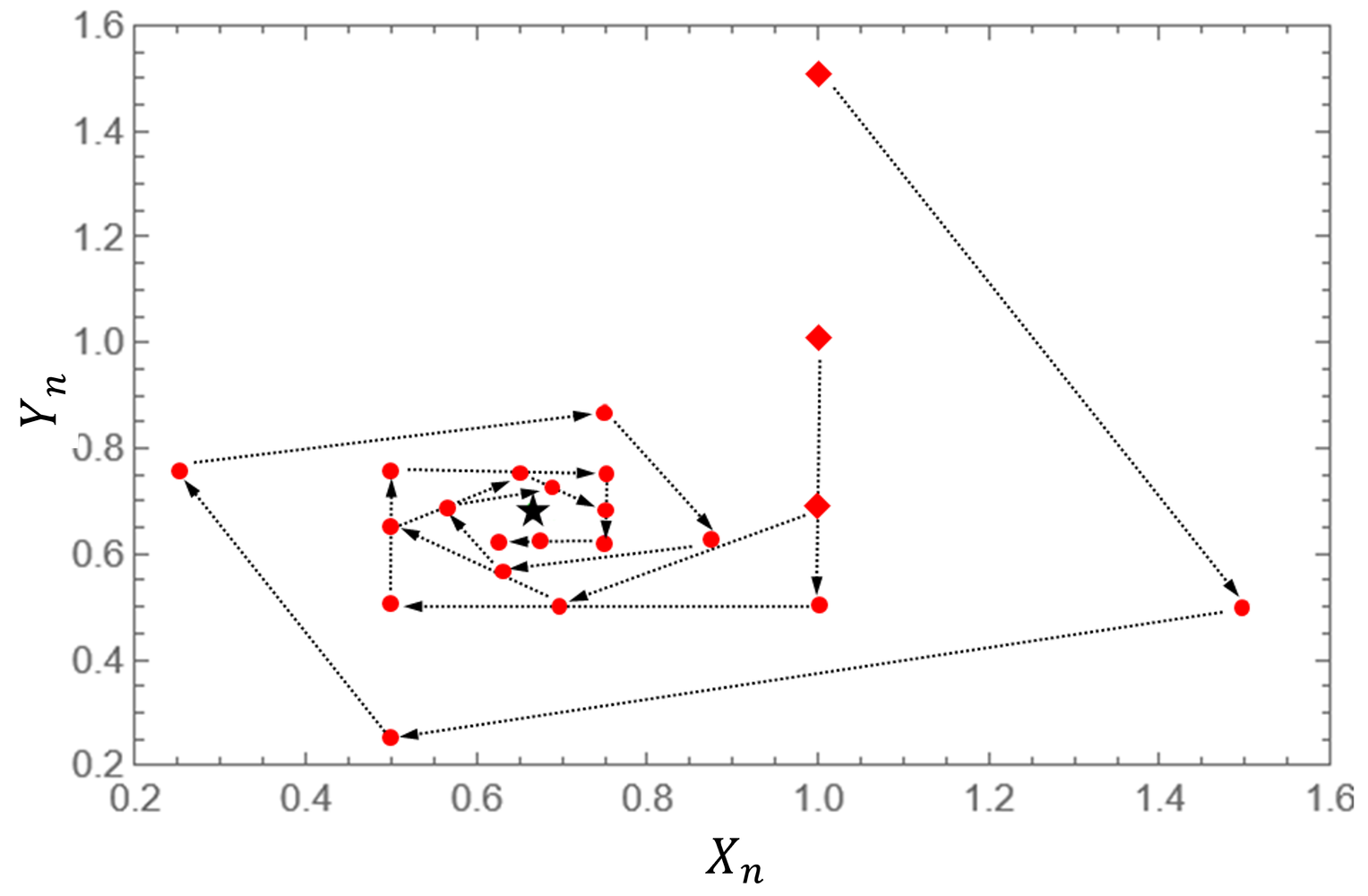}
        \\
        \hspace{0.3cm}
        (a)
        \hspace{3.5cm}
        (b)\\
        \caption{\label{Fig.m<=1} 
            Trajectories starting 
            from three different filled squares for 
            (a) $m=1$ and (b) $m=0.5$. 
            Now we set $B=1$.
            The star in each figure shows the fixed point $\bm{\bar x}_{I} = \left(\frac{B}{1+m},\frac{B}{1+m}\right)$.
            The four green open circles in (a) show the ultradiscrete states in $\mathcal{C}(m=1)$.
        }
    \end{center}
\end{figure}

To grasp the dynamical properties of  $\mathcal{C}$ and $\mathcal{C}_s$, 
we introduce the Poincar\'{e} section $L \equiv \{(X, B), X > 0\}$
\cite{Ohmori2022a}.
Note that every trajectory possesses a point on the line $L$.
In particular, $\bm{x}_{0}^{\mathcal{C}}$ and $\bm{x}_{0}^{\mathcal{C}_s}$ are on $L$ and return to themselves.
The Poincar\'{e} map $P_m$ on $L$ is constructed by considering the next return point $X_{n+1}$ on $L$ for the trajectory 
from the point $X_{n}$ on $L$.
Actually when $m > 1$, $P_m$ is obtained as the one-dimensional piecewise linear discrete dynamical system, $X_{n+1} = P_m(X_n)$, 
where 
\begin{align}
    P_m(X_n)=	
    \begin{cases}
        B   &  \left( 0<X_n \leq \frac{(m-1)B}{m^2} \right ),  \\
        m^2X_n+(1-m)B  &  \left( \frac{(m-1)B}{m^2}<X_n < \frac{B}{m} \right ),  \\
        B & \left( \frac{B}{m} \leq X_n \right ).
    \end{cases}
    \label{eqn:Pmap}
\end{align}
Figure \ref{Fig.P_map} (a) shows the graph of $X_{n+1}=P_m(X_n)$ for $m>1$.
This graph intersects the line $X_{n+1}=X_{n}$ 
at the two points $X_n=X_s\equiv \frac{B}{m+1}$ and $X_n=B$,
which are found to be unstable and stable, respectively.
Therefore, we conclude that $\mathcal {C}$ ($\mathcal {C}_s$) is an attracting (repelling) limit cycle.
%
%
From the graph of $P_m$, it is also found that the slope of $P_m$ tends to $1$ when $m \to 1$ as shown in Fig. \ref{Fig.P_map} (b).
Then, a trajectory starting from a point outside of $\mathcal{C}(m=1)$ converges
to $\mathcal{C}(m=1)$.
On the other hand,  
a trajectory starting from inside of $\mathcal{C}(m=1)$ becomes a different cycle dependent on the initial states 
around the fixed point $\left(\frac{B}{2},\frac{B}{2}\right)$ 
as shown in Fig. \ref{Fig.m<=1} (a).
Therefore, $\mathcal{C}(m=1)$ is the half-stable limit cycle.  

When $m<1$, the time evolution of $Y_{n}$ for eq.(\ref{eqn:1-1a}) 
can be written as $Y_{n+2}=B-\max(0, m Y_{n})$.
It is found from this time evolution that any initial state 
finally converges to $\left(\frac{B}{m+1},\frac{B}{m+1}\right)$, 
which is the spiral sink.
Therefore, in eq.(\ref{eqn:1-1a}) with $B>0$, $m = 1$ becomes the bifurcation point for the Neimark-Sacker bifurcation and 
the two limit cycles $\mathcal{C}$ and $\mathcal{C}_s$ 
emerge when $B>0$ and $m>1$.

\begin{figure}[t!]
    \begin{center}
    \includegraphics[width=7cm]{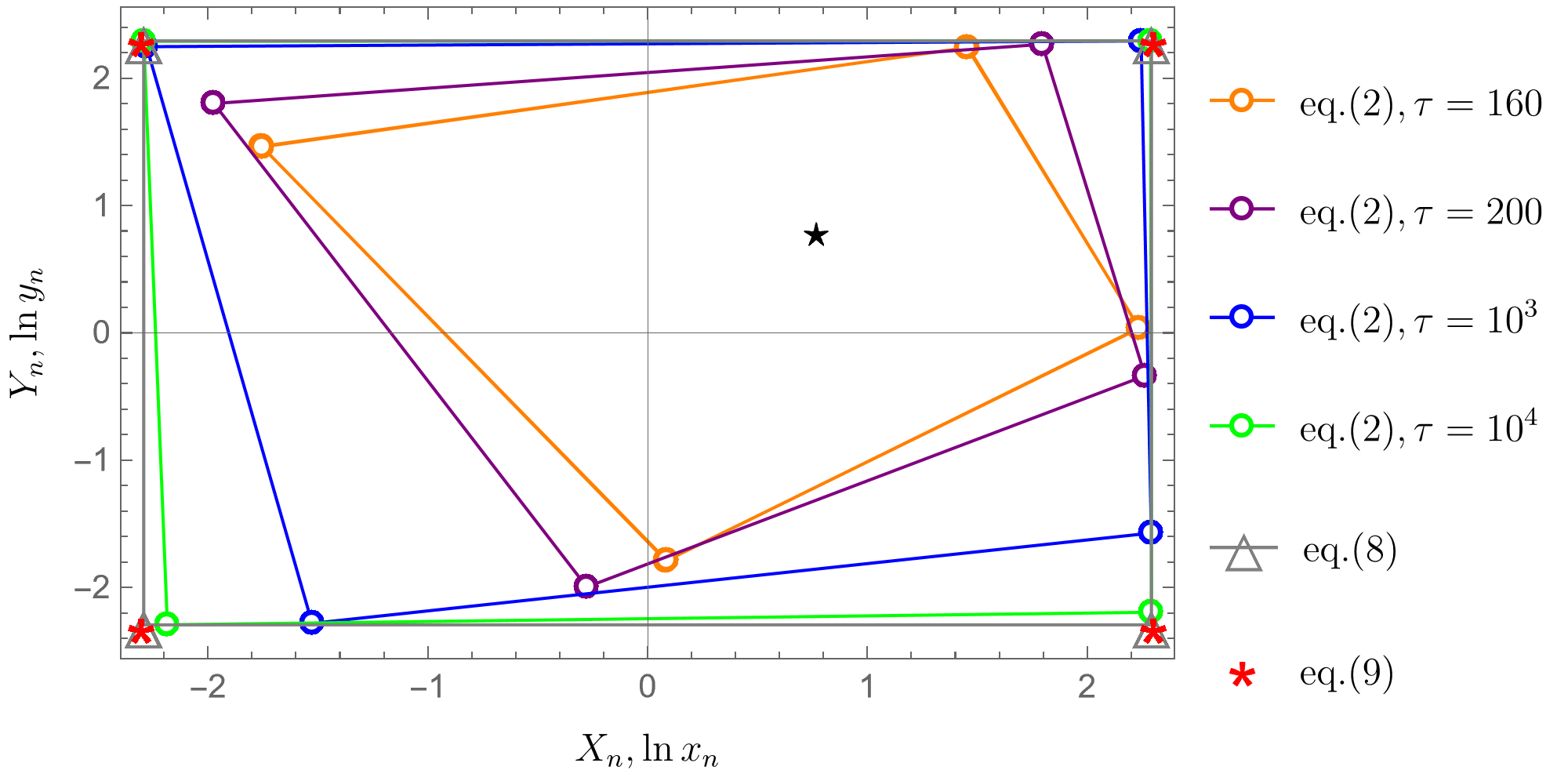}
    \\
    \caption{\label{Transition_to_UDD}
    Circles show limit cycle states with four different 
    values of $\tau$ in eq.(\ref{eqn:trop1}). 
    Triangles show the case of eq.(\ref{eqn:trop1_inf}).
    Red asterisks and the black star show the limit cycle states and the fixed point obtained from the max-plus equations, respectively.
    We set $b=10$ ($B = \ln b$).}
    \end{center}
\end{figure}

Finally, we briefly comment on relationship between limit cycle solutions 
of eqs.(\ref{eqn:trop1}), (\ref{eqn:trop1_inf}) and that of eq.(\ref{eqn:1-1a}).
Figure \ref{Transition_to_UDD} shows comparison of the four limit cycles with different values 
of $\tau$ in eq.(\ref{eqn:trop1}) 
and the limit cycle of eq.(\ref{eqn:trop1_inf}) 
with that of the max-plus equation, eq.(\ref{eqn:1-1a}).
It is found that as $\tau$ increases the limit cycle states of the tropically 
discretized equation approach those of the max-plus equation.
%

\section{Summary and Conclusion}

We have investigated the dynamical properties of 
the tropically discretized and the max-plus negative feedback models.
For the tropically discretized model, we analytically identify conditions under which the Neimark-Sacker bifurcation occurs and the limit cycle soluions emerge, in a systematic manner.
We find the ultradiscrete limit cycle with four states emerges 
when $\tau$ is large even for $\tau \to \infty$.
Furthermore for the ultradiscrete max-plus model, 
the two limit cycles, $\mathcal {C}$ and $\mathcal {C}_s$, 
emerge when $B>0$ and $m>1$.
These limit cycles have been analyzed by using the Poincar\'{e} map method, 
and we find that $\mathcal {C}$ is stable and $\mathcal {C}_s$ is unstable.
We have also confirmed that the limit cycle solutions by the tropically discretized model become close to those by the max-plus model when $\tau \to \infty$.
The dynamical behavior of the limit cycles 
for the tropically discretized equations as $\tau$ increases and the approach to the limit cycle for the max-plus model when $\tau$ tends to infinity are also observed 
in Sel'kov model\cite{Ohmori2022a,Yamazaki2021}, 
suggesting that they are general characteristics.
%
%


\bigskip

\noindent
{\bf acknowledgments}

The authors are grateful to 
Prof. M. Murata, Prof. K. Matsuya, 
Prof. D. Takahashi, 
Prof. T. Yamamoto, and Prof. Emeritus A. Kitada 
for useful comments and encouragements. 
This work was supported by JSPS
KAKENHI Grant Numbers 22K13963 and 22K03442.


%

\end{document}